# Simulating the Evolution of Signaling Signatures during CART-Cell – Tumor Cell Interactions

Viren Shah, Justin Womack, Anthony E. Zamora, Scott S. Terhune, and Ranjan K. Dash, *Member, IEEE EMBS*

*Abstract*– Immunotherapies have been proven to have significant therapeutic efficacy in the treatment of cancer. The last decade has seen adoptive cell therapies, such as chimeric antigen receptor T-cell (CART-cell) therapy, gain FDA approval against specific cancers. Additionally, there are numerous clinical trials ongoing investigating additional designs and targets. Nevertheless, despite the excitement and promising potential of CART-cell therapy, response rates to therapy vary greatly between studies, patients, and cancers. There remains an unmet need to develop computational frameworks that more accurately predict CART-cell function and clinical efficacy. Here we present a coarse-grained model simulated with logical rules that demonstrates the evolution of signaling signatures following the interaction between CART-cells and tumor cells and allows for *in silico* based prediction of CART-cell functionality prior to experimentation.

*Clinical Relevance*– Analysis of CART-cell signaling signatures can inform future CAR receptor design and combination therapy approaches aimed at improving therapy response.

*Keywords*– CAR T-cell, CAR-T therapy, Computational model, Mathematical model, Boolean model, Logic-based model.

## I. INTRODUCTION

Chimeric antigen receptor T-cell (CART-cell) therapy is an adoptive cell therapy in which a patient's T-cells are genetically modified to express chimeric antigen receptor (CAR) constructs. The CAR is a fusion of protein domains consisting of extracellular antigen targeting and intracellular signaling and costimulatory domains. The targeting domains are antibody-derived single chain variable fragments (scFv). The signaling domain is T-cell receptor (TCR)-derived CD3ζ with CD28 or 4-1BB typically used as costimulatory domains [1]. This construct allows CART-cells to target and eliminate antigen-expressing cancer cells.

Therapy dynamics involve interconnected activation, signaling, inhibitive, cytotoxic pathways including crosstalk and feedback between and within effector and target cells. Differences in any one variable (molecular or environmental) can lead to non-intuitive effects on CART function and impact therapy dynamics. It is not feasible to test clinically or experimentally the vast range of perturbations which may influence therapy responses, but these perturbations can offer significant insight on how CART products may be improved.

Computational modeling of molecular signaling networks is a powerful tool for understanding the dynamics of complex systems. These models can generate testable hypotheses and predict different possible non-intuitive outcomes of the system. Creating clinically valuable differential equations-based models requires gathering highly specific and robust datasets which are currently experimentally cost-prohibitive. In contrast, logic-based approaches allow the creation of networks based on understanding of biological functions (e.g., activation, inhibition, etc.) of individual components to study the behavior of the system under different perturbations.

While literature on CART-cell computational modeling is sparse, logic models of T-cell activation, cytokine and inhibitory signaling have been previously investigated [2-4]. Motivated by these approaches, here we define a coarse-grained logic-based model that can be used to investigate the dynamic evolution of the complex regulatory signaling network governing interactions between CART-cells and tumor cells. The model is used to explore how changes in signaling pathways impact signaling signatures indicative of specific functional states of the interaction network.

## II. METHODS

### A. Signaling Network

A minimal model (**Fig. 1A**) of the signaling network of CART-cells (consisting of CD3ζ activating and CD28 costimulatory domains) during interactions with targeted tumor cells was derived using curated pathway diagrams for T-cell activation and literature sources for transcriptional and gene regulation. Both literature sources of T-cell signaling and CART-cell signaling are considered as networks overlap [5, 6]. Exceptions arise in CD28 and LAT signal transduction [7], which are incorporated. The constructed network diagrams the signaling activity and cellular responses following engagement of tumor cells by CART-cells. Major activation (CAR, MAP kinase (MAPK), PI3K-AKT-MTOR1, calcium, and other signaling intermediates), cytokine (IL2), cytotoxic, and inhibitory (PD1 and CTLA4) signaling pathways leading to transcription regulation and gene expression which influence CART-cell function are considered. Relevant crosstalks, feedforward and feedback mechanisms are also considered. For simplicity, pathway intermediates which help propagate signaling activity but were not points of pathway branching, regulatory activity, or feedbacks and feedforwards were removed from the constructed network. Additionally, pathway intermediates which act in concert were combined.

*Activation:* The CAR considered for the signaling network consists of CD3ζ activating and CD28 costimulatory domains. Ligation of the CAR by tumor antigens (TA) through an active immune synapse (IS) leads to LCK facilitated activation of CD28, CD3ζ, and ZAP70 [8]. Basal pools of active LCK are constitutively maintained in T-cells to initiate signaling once engaged by targets [9]. Activated CD28 initiates activation of PI3K-AKT-MTORC1 [6]. Activated ZAP70 mediates signal transduction into calcium, MAPK, and PKC pathways through various intermediates leading to the activation of transcription factor (TF) CFOS, CJUN, NFκB, NFAT, and CREB. TF CJUN and CFOS combine to form AP1 [6]. AP1, NFAT and NFκB are required for IL2 gene expression

*This work was partly supported by NIH grant R21-AI149039 to S.S.T. and R.K.D. The funders had no role in study design, data collection and analysis, decision to publish, or preparation of the manuscript.

All authors are with the Medical College of Wisconsin (virshah@mcw.edu, jwomack@mcw.edu, azamora@mcw.edu, sterhune@mcw.edu, rdash@mcw.edu; (414) 955-4497).

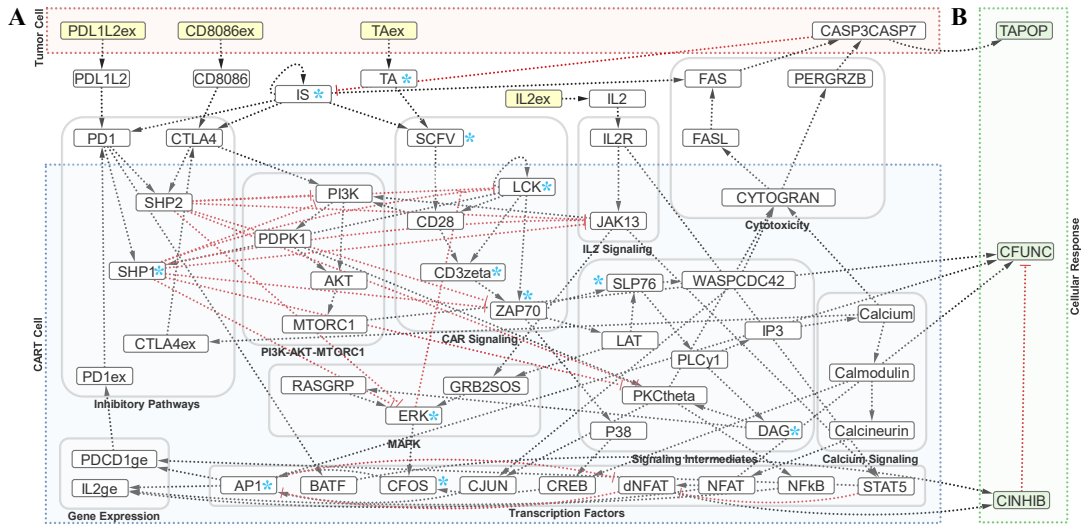

Fig. 1. Diagram of interaction network regulating CART-cell function. (A) CART and tumor cell compartments are boxed along with notable signaling pathways. Nodes: Input nodes are shaded yellow, internal nodes are shaded white, (B) cellular response nodes for CART-cell function (CFUNC), CART-cell inhibition (CINHIB) and tumor apoptosis (TAPOP) are green. Edges: Activating edges are shown in black, inhibitory edges are red, rule sets for multiple edges further specified in Table 1. Selected nodes for Fig. 2 are marked with blue *. Abbreviations: ex: expression, TA: tumor antigen, ge: gene expression.

(IL2ge) [10], while AP1, NFAT also promote *pdcd1* gene expression (PDCD1ge) [11]. In the absence of AP1, NFAT dimerizes and promotes anergy and exhaustion in T-cells [12].

*Cytotoxicity:* Cytotoxicity in T-cells is mediated through the release of perforin/granzyme B (PERGRZB) and surface molecules such as Fas ligand (FASL) from cytotoxic granules (CYTOGRAN). This process requires actin cytoskeleton reorganization supported by WASP and cdc42 (WASPCDC42), polarizing microtubules organizing centers toward the IS (PKCθ), and granule secretion (supported by calcium and PI3K) [13]. FASL signals through the IS to its receptor FAS, and both FAS and PERGRZB initiate apoptotic signaling cascades leading to caspases 3 and 7 (CASP3CASP7) [6].

*Cytokine:* IL2 signals through IL2R leading to the activation of kinases JAK1 and JAK3 (JAK13), which together activate PI3K-AKT-MTORC1, and MAPK pathways. IL2R, and JAK13 activate TF STAT5. STAT5 and NFAT promote *PDCD1ge* [10]. For simplicity receptor affinity is not distinguished, which assumes availability of high affinity IL2R.

*Inhibitory Signaling:* Inhibitory signaling is mediated by engagement of ligands PDL1 or PDL2 (PDL1L2) with PD1 and CD80 or CD86 (CD8086) with CTLA4 through an active IS. Engagement of these receptors activates the phosphatase SHP2 which counteracts the kinase activity of LCK, ZAP70, PKCθ, JAK1, JAK3, PI3K, and ERK. In the absence of SHP2, PD1 is also able to signal through SHP1 which has similar activity [14]. PD1 is expressed through regulation of PDCD1ge [11], and CTLA4 is expressed by secretion from intracellular granules (supported by calcium) following T-cell activation [15]. Activation of PD1 induces the TF BAFT which promotes exhaustion in T-cells [16].

*Crosstalks, feedforward, and feedback loops:* Activated LCK and CD3ζ initiate negative feedback of LCK to dampen activation signaling through SHP1 [8]. Activated ERK phosphorylates LCK preventing interactions with SHP1 and SHP2, and, separately, activation of SHP1 by LCK [9].

The complied network is a coarse-grained curated construction of the biological network which can be modularly extended to include additional factors and pathways contributing to network regulation not considered here.

### B. Model and simulation

The rules governing the nodes and edges of the constructed model and corresponding literature references are detailed in **Table 1**. The logic model is designed and simulated using the stochastic discrete logic-based formalism described in [17] for modeling probability Boolean networks (PBNs). Each node in the model is defined by a logical rule (update function), which when satisfied switches (or keeps) the node ON (value of 1) or otherwise switches (or keeps) the node OFF (value of 0). The update function is supported by corresponding propensity probabilities, $p_a$ and $p_d$, which respectively set the probability that a given update function will update during a given iteration based on its current state (either ON or OFF). This is done to account for inherent randomness in the system where not all process (reactions or sets of reactions) can be assumed to occur during each iteration. For similar approaches to dynamic modeling of Boolean and logic based networks, see [18], and [19].

Each update function considers edges (connections between nodes) which participate in processes to turn on a node (e.g., phosphorylation by a kinase) or inhibit these processes (e.g., dephosphorylation by a phosphatase). Default probabilities for each update function are set to $p_a = 0.5$. In the absence of continuous activating signals, activated signaling proteins and molecules in cells are degraded through a variety of processes (e.g., ubiquitination). To account for this each update function is assigned a $p_d = 0.05$. High probabilities of node activation alongside low probabilities of node degradation are assumed to match kinetic data on T-cell activation responses showing bimodal all-or-none responses and degradation processes slowly attenuating these signals thereafter [20].

Three readout nodes are included which capture the trajectories of signaling signatures **(Fig. 1B)**. The CART-cell function (CFUNC) node captures activation and proliferation signals by proxy of AP1, NFAT, and MTORC1. The CART-cell inhibition node (CINHB) consolidates signals leading to anergic and exhausted states by proxy of dNFAT and BATF. The tumor apoptosis node (TAPOP) captures activity of cytotoxic pathways leading to apoptosis through CASP3CASP7.

TABLE I.    MODEL NODE DEFINITIONS

| # | node | update function[a] | ref. |
|---|---|---|---|
| 1 | TAex | input | - |
| 2 | TA | TAex | - |
| 3 | IS | !CASP3CASP7 | - |
| 4 | SCFV | IS & TA | [21] |
| 5 | CD28 | SCFV & LCK | [21] |
| 6 | CD3zeta | LCK & CD28 | [21] |
| 7 | LCK | !LCK & ((!SHP1 | (SHP1 & ERK)) & ((!SHP2 | (SHP2 & ERK)) | [8, 9] |
| 8 | ZAP70 | CD3zeta & LCK & !(SHP1 | SHP2) | [6, 21] |
| 9 | IL2ex | input | - |
| 10 | IL2 | IL2ex | [10] |
| 11 | IL2R | IL2 | [10] |
| 12 | JAK13 | IL2R | [10] |
| 13 | PDL1L2ex | input | - |
| 14 | PDL1L2 | PDL1L2ex | [6] |
| 15 | PD1ex | PDCD1ge | [11] |
| 16 | PD1 | IS & PDL1L2 | [6] |
| 17 | CD8086ex | input | - |
| 18 | CD8086 | CD8086ex | [6] |
| 19 | CTLA4ex | Calcium | [15] |
| 20 | CTLA4 | IS & CD8086 & CTLA4ex | [6] |
| 21 | SHP2 | PD1 | CTLA4 | [14, 15] |
| 22 | SHP1 | (PD1 & !SHP2) | (LCK & !ERK) | [9, 14] |
| 23 | LAT | ZAP70 | [5, 6] |
| 24 | SLP76 | ZAP70 | LAT | [5, 6] |
| 25 | PLCgamma1 | SLP76 | [5, 6] |
| 26 | DAG | PLCgamma1 | [5, 6] |
| 27 | IP3 | PLCgamma1 | [5, 6] |
| 28 | WASPCDC42 | SLP76 | [5, 6] |
| 29 | P38 | ZAP70 | [5, 6] |
| 30 | PKCtheta | (DAG | PDPK1) & !(SHP1 | SHP2) | [5, 6] |
| 31 | PI3K | CD28 | CTLA4 | [5, 6] |
| 32 | PDPK1 | PI3K | [5, 6] |
| 33 | AKT | PI3K & PDPK1 | [5, 6] |
| 34 | MTORC1 | AKT | [5, 6] |
| 35 | RASGRP | DAG | [5, 6] |
| 36 | GRB2SOS | LAT | JAK13 | CD28 | [5, 6] |
| 37 | ERK | (RASGRP | GRB2SOS) & !(SHP1 | SHP2) | [5, 6] |
| 38 | Calcium | IP3 | [5, 6] |
| 39 | Calmodulin | Calcium | [5, 6] |
| 40 | Calcineurin | Calcium & Calmodulin | [5, 6] |
| 41 | CFOS | ERK | CREB | [5, 6] |
| 42 | CJUN | PKCtheta | WASPCDC42 | [5, 6] |
| 43 | AP1 | CFOS & CJUN & !dNFAT | [5, 6] |
| 44 | BATF | PD1 | [16] |
| 45 | NFKB | PKCtheta | [5, 6] |
| 46 | NFAT | Calcineurin | P38 | [5, 6] |
| 47 | dNFAT | (NFAT & !AP1) | (NFAT & !STAT5) | [12] |
| 48 | CREB | P38 | Calmodulin | [6] |
| 49 | STAT5 | IL2R & JAK13 | [10] |
| 50 | IL2ge | AP1 & NFKB & NFAT | [10] |
| 51 | PDCD1ge | (AP1 & NFAT) | (STAT5 & NFAT) | [11] |
| 52 | CYTOGRAN | WASPCDC42 & PKCtheta & Calcium & PI3K | [13] |
| 53 | FASL | CYTOGRAN | [6] |
| 54 | PERGRZM | CYTOGRAN | [6] |
| 55 | FAS | FASL | [6] |
| 56 | CASP3CASP7 | FASL | PERGRZB | [6] |
| 57 | CFUNC | AP1 & NFAT & MTORC1 & !CDSYF | - |
| 58 | CINHI | dNFAT | BATF | - |
| 59 | TAPOP | CASP3CASP7 | - |

a. Logical operator symbols: & (AND), | (OR), ! (NOT)

The simulation is cast as an enclosed loop with the intention to model the evolution of signaling signatures through the continuous activation of CART-cells by interactions with tumor cells. This is intended to be representative of the initial evolution of signaling events in CART-cell therapy.

A random order asynchronous update scheme is used for all nodes. This assumes all nodes are updated rapidly but may update in varying order and leads to stochastic (as opposed to deterministic) trajectories which is better representative of the heterogenous biological system that consists of multiple CART-cell types (e.g., CD4 vs. CD8) and different states of differentiation. The model does not explicitly model or differentiate between these states.

All results are generated from averaging 10,000 simulations. Initial states of input nodes are randomized when activated. Input nodes are held at a constant 0 when inactive. Initial states of all other nodes set to 0 except LCK which is assumed to be active at basal states [8]. The model was simulated using MATLAB R2022A. Simulation code is available on GitHub: github.com/MCWComputationalBiologyLab/Shah_2023_IEEE.

### III. RESULTS

We first consider an example of the dynamic behavior of the network following engagement of the CAR (**Fig. 2**). The selected nodes follow the marked pathway in **Fig. 1A**. Engagement of the CAR transduces activating signals from the receptor to TF AP1. The y-axis (activity level) indicates the average number of nodes in an ON state during a given iteration. This can be interpreted as the portion of the population exhibiting active signaling through that node, however this does not necessarily correlate with absolute biological concentrations. The x-axis (time in arbitrary units (a.u.)) indicates the trajectory of the average response at each node which correlates with biological sequences but not real time. The present results for ERK activation roughly match experimental kinetics [22] and the AP1 response is consistent with a bimodal response to activation signals [9].

We next investigated how ligation of different receptors in the interaction network changed the trajectories and steady-state behavior of signaling signatures. Activation of only the CAR receptor (**Fig. 3A**) leads to high-moderate functional and cytotoxic signaling signatures. The y-axis here is interpreted as the portion of the population that exhibits the CFUNC ($p_{cfunc}$), CINHIB ($p_{cihib}$), and TAPOP ($p_{tapop}$) signaling signatures. Addition of IL2 leads to increased functional signaling and comparable cytotoxic signaling (**Fig. 3B**). Making inhibitory pathways through PDL1L2 and CD8086 active leads to a marked decrease in functional signaling signatures and reducing cytotoxic signatures to a lesser degree (**Fig. 3C**). The simulation indicates that in the presence of inhibitory signals the network activates functional signaling initially, but this trajectory is dampened by inhibitory feedback loops. This trajectory is likely representative of the *in vivo* system. It is important to note that in this model inhibitory signaling is transient as we do not model terminal differentiation, and as such there is continuous competition between activating and inhibitory signals leading to the observed behavior. Interestingly only blocking signaling through PDL1L2 leads to an increase in functional signaling, while CD8086 blockade has negligible impact (**Figs. 3D,E**). This is supported by experiments showing CART-cells with CD28 intercellular domains have attenuated sensitivity to CTL4 inhibition [23]. Possible mechanisms are elimination of ligation competition and compensation through PD1 pathways which both activate SHP2.

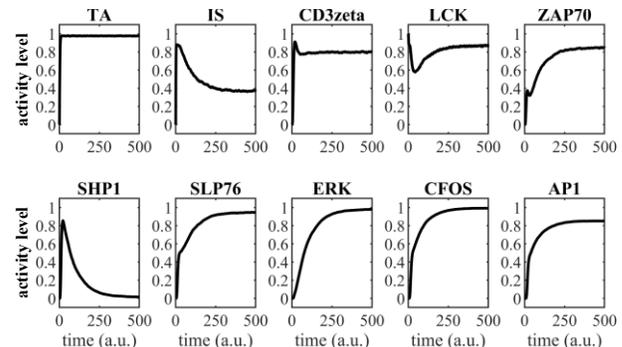

Fig. 2: Model simulation of signal propagation through the network. Abbreviations: a.u.: arbitrary units.

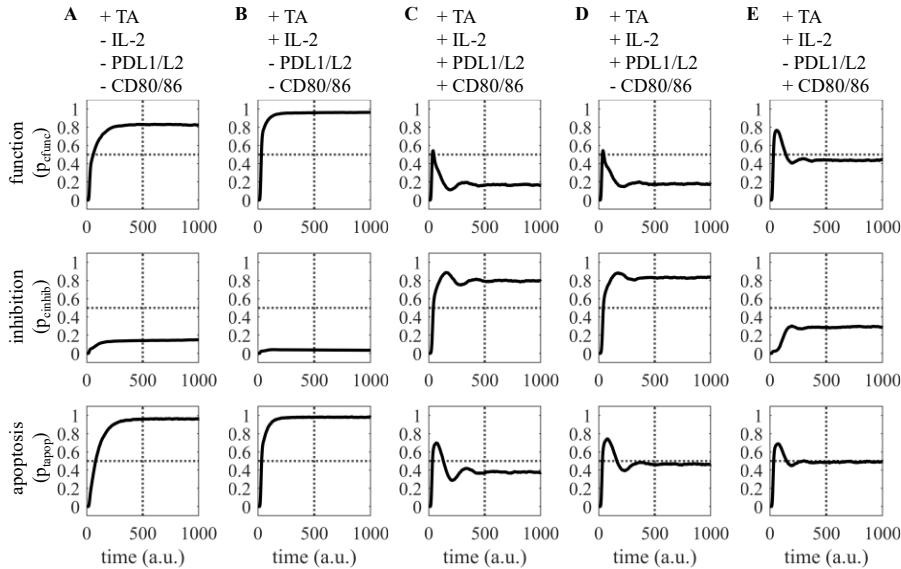

Fig. 3: Model response to receptor ligation. (A) Portion of population displaying signatures of CART function ($p_{cfunc}$), CART inhibition ($p_{cinhib}$), and tumor apoptosis ($pt_{apop}$) with CAR ligation. (B) Responses upon adding IL2, C) PDL1L2 and CD8086, or D) only PDL1L2 or E) only CD8086 ($p_{tapop}$ = portion of tumor apoptosis; $p_{cinhib}$ = portion of CART cellular inhibition; $p_{cfunc}$ = portion of CART cellular function). Abbreviations: a.u.: arbitrary units.

Using **Fig. 3C** steady state results as the control, which represents the scenario where signaling is possible through all receptor interactions, we perturbed the network at individual nodes to determine how the steady state signaling signatures of the network changed in response (**Fig. 4**). The model re-states expected behavior where deficiencies in critical activation nodes lead to curtailed ability to maintain functional signaling, whilst overexpression of these pathways induces inhibition, which aligns with experimental findings [24]. Upregulation of intermediate activation signaling molecules are shown to have slight propensity to increase functional signaling (PI3K, ERK, PKC), while abrogation of inhibitory pathways and downstream TF (PD1, BATF) significantly improve function but not tumor apoptosis signatures. The model fails to respond to experimental findings such as CJUN overexpression curtailing exhaustion signatures [25], which indicates that additional transcription regulation not considered in this model is present. Including additional TF and gene regulation will allow more refined definition of feedback mechanisms and more robust analysis of network perturbations.

## IV. DISCUSSION

Our aim with this work was to investigate the feasibility of simulating the dynamic evolution of signaling signatures following interactions between CART-cells and tumor cells without explicit knowledge of parameters and mechanisms which influence signaling events. The utilized method presents a simplistic approach that can be used to consolidate current understanding of signaling pathways and perform *in silico* network perturbations.

Beyond PD1 inhibition, disrupting additional central signaling nodes generally resulted in loss of function, as we have defined it here. In contrast, we failed to observe substantial increases in CART-cell activities upon simulating overexpression of activating nodes. Therefore, the model simulations predict that significant improvement to baseline functional and cytotoxic ability by perturbing the intercellular network are possibly intrinsically limited. While the activity of inhibitory pathways dampens functional signaling, the ability of the interaction network to promote apoptosis with a non-negligible probability is maintained. In the context of response to CART-cell therapy, this indicates that disrupted intercellular signaling instead of increased inhibitory signaling are more likely to abrogate function. Consequently, optimization strategies focused on supporting low steady-state activity of the activation network are potentially more likely to promote therapy response. It should be noted that both differential event sequences (e.g., prior IL2 priming) and additional signaling through modeled pathways (e.g., simultaneous signaling through CAR and TCR) would both impact the presented model dynamics and need to be further investigated.

The presented approach is limited to demonstrating how signaling dynamics in CART cells may sequentially evolve. Transitioning the model to a semi-quantitative approach that links signal transduction events with proliferative and apoptotic events on appropriate time scales is a reasonable next step. Supported with data from single cell *in vitro* assays of CART and tumor cell interactions to define necessary parameters, this approach would facilitate analysis of CART and tumor cell population dynamics in continuous time.

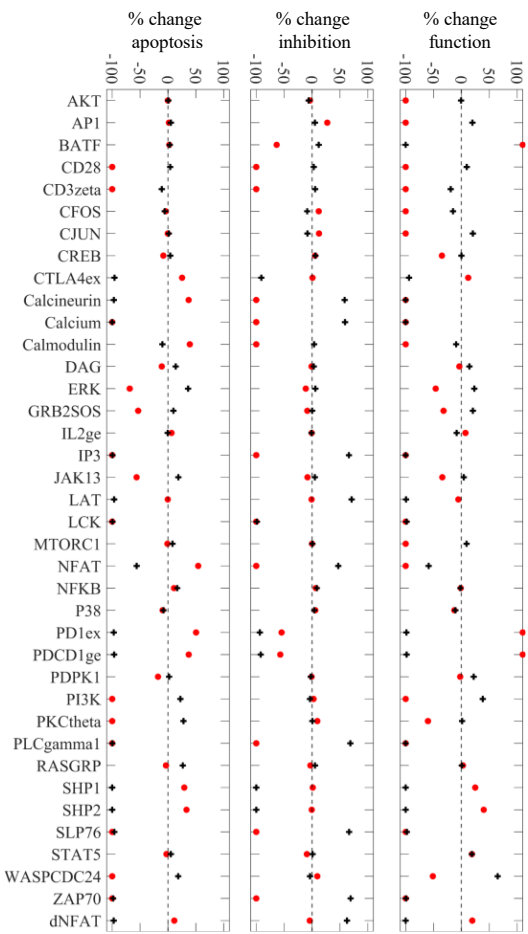

Fig. 4. Functional outcomes upon perturbing internal CART-cell network nodes. Percent change in the portion of signaling signatures for CART-cell function ($p_{cfunc}$), CART-cell inhibition ($p_{cinhib}$), and tumor apoptosis ($p_{tapop}$) upon forcing nodes constitutively off (I.C. = 0, $p_a$ = 0, red *) to simulate knock-out or constitutively on (I.C. = 1, $p_a$ = 1, black +) to simulate overexpression of internal CART-cell signaling nodes. On plot line > |±100|.

## REFERENCES

1  Dotti, G., Gottschalk, S., Savoldo, B., and Brenner, M.K.: 'Design and development of therapies using chimeric antigen receptor-expressing T cells', Immunological Reviews, 2014, 257, (1), pp. 107-126
2  Saez-Rodriguez, J., Simeoni, L., Lindquist, J.A., Hemenway, R., Bommhardt, U., Arndt, B., Haus, U.-U., Weismantel, R., Gilles, E.D., Klamt, S., and Schraven, B.: 'A Logical Model Provides Insights into T Cell Receptor Signaling', PLoS Computational Biology, 2007, 3, (8)
3  Hernandez, C., Thomas-Chollier, M., Naldi, A., and Thieffry, D.: 'Computational Verification of Large Logical Models—Application to the Prediction of T Cell Response to Checkpoint Inhibitors', Frontiers in Physiology, 2020, 11
4  Thieffry, D., Beyer, T., Busse, M., Hristov, K., Gurbiel, S., Smida, M., Haus, U.-U., Ballerstein, K., Pfeuffer, F., Weismantel, R., Schraven, B., and Lindquist, J.A.: 'Integrating Signals from the T-Cell Receptor and the Interleukin-2 Receptor', PLoS Computational Biology, 2011, 7, (8)
5  Hwang, J.-R., Byeon, Y., Kim, D., and Park, S.-G.: 'Recent insights of T cell receptor-mediated signaling pathways for T cell activation and development', Experimental & Molecular Medicine, 2020, 52, (5), pp. 750-761
6  Kanehisa, M., and Goto, S.: 'KEGG: Kyoto Encyclopedia of Genes and Genomes', Nucleic Acids Research, 2000, 28, (1), pp. 27-30
7  Dong, R., Libby, K.A., Blaeschke, F., Fuchs, W., Marson, A., Vale, R.D., and Su, X.: 'Rewired signaling network in T cells expressing the chimeric antigen receptor (CAR)', The EMBO Journal, 2020, 39, (16)
8  Nika, K., Soldani, C., Salek, M., Paster, W., Gray, A., Etzensperger, R., Fugger, L., Polzella, P., Cerundolo, V., Dushek, O., Höfer, T., Viola, A., and Acuto, O.: 'Constitutively Active Lck Kinase in T Cells Drives Antigen Receptor Signal Transduction', Immunity, 2010, 32, (6), pp. 766-777
9  Štefanová, I., Hemmer, B., Vergelli, M., Martin, R., Biddison, W.E., and Germain, R.N.: 'TCR ligand discrimination is enforced by competing ERK positive and SHP-1 negative feedback pathways', Nature Immunology, 2003, 4, (3), pp. 248-254
10  Liao, W., Lin, J.-X., and Leonard, Warren J.: 'Interleukin-2 at the Crossroads of Effector Responses, Tolerance, and Immunotherapy', Immunity, 2013, 38, (1), pp. 13-25
11  Bally, A.P.R., Austin, J.W., and Boss, J.M.: 'Genetic and Epigenetic Regulation of PD-1 Expression', The Journal of Immunology, 2016, 196, (6), pp. 2431-2437
12  Mognol, G.P., González-Avalos, E., Ghosh, S., Spreafico, R., Gudlur, A., Rao, A., Damoiseaux, R., and Hogan, P.G.: 'Targeting the NFAT:AP-1 transcriptional complex on DNA with a small-molecule inhibitor', Proceedings of the National Academy of Sciences, 2019, 116, (20), pp. 9959-9968
13  Kabanova, A., Zurli, V., and Baldari, C.T.: 'Signals Controlling Lytic Granule Polarization at the Cytotoxic Immune Synapse', Frontiers in Immunology, 2018, 9
14  Celis-Gutierrez, J., Blattmann, P., Zhai, Y., Jarmuzynski, N., Ruminski, K., Grégoire, C., Ounoughene, Y., Fiore, F., Aebersold, R., Roncagalli, R., Gstaiger, M., and Malissen, B.: 'Quantitative Interactomics in Primary T Cells Provides a Rationale for Concomitant PD-1 and BTLA Coinhibitor Blockade in Cancer Immunotherapy', Cell Reports, 2019, 27, (11), pp. 3315-3330.e3317
15  Rudd, C.E., Taylor, A., and Schneider, H.: 'CD28 and CTLA-4 coreceptor expression and signal transduction', Immunological Reviews, 2009, 229, (1), pp. 12-26
16  Quigley, M., Pereyra, F., Nilsson, B., Porichis, F., Fonseca, C., Eichbaum, Q., Julg, B., Jesneck, J.L., Brosnahan, K., Imam, S., Russell, K., Toth, I., Piechocka-Trocha, A., Dolfi, D., Angelosanto, J., Crawford, A., Shin, H., Kwon, D.S., Zupkosky, J., Francisco, L., Freeman, G.J., Wherry, E.J., Kaufmann, D.E., Walker, B.D., Ebert, B., and Haining, W.N.: 'Transcriptional analysis of HIV-specific CD8+ T cells shows that PD-1 inhibits T cell function by upregulating BATF', Nature Medicine, 2010, 16, (10), pp. 1147-1151
17  Murrugarra, D., Veliz-Cuba, A., Aguilar, B., Arat, S., and Laubenbacher, R.: 'Modeling stochasticity and variability in gene regulatory networks', EURASIP Journal on Bioinformatics and Systems Biology, 2012, 2012, (1)
18  Konstorum, A., Vella, A.T., Adler, A.J., and Laubenbacher, R.C.: 'A mathematical model of combined CD8 T cell costimulation by 4-1BB (CD137) and OX40 (CD134) receptors', Scientific Reports, 2019, 9, (1)
19  Stoll, G., Viara, E., Barillot, E., and Calzone, L.: 'Continuous time boolean modeling for biological signaling: application of Gillespie algorithm', BMC Systems Biology, 2012, 6, (1)
20  Lee, K.-H., Dinner, A.R., Tu, C., Campi, G., Raychaudhuri, S., Varma, R., Sims, T.N., Burack, W.R., Wu, H., Wang, J., Kanagawa, O., Markiewicz, M., Allen, P.M., Dustin, M.L., Chakraborty, A.K., and Shaw, A.S.: 'The Immunological Synapse Balances T Cell Receptor Signaling and Degradation', Science, 2003, 302, (5648), pp. 1218-1222
21  Ramello, M.C., Benzaïd, I., Kuenzi, B.M., Lienlaf-Moreno, M., Kandell, W.M., Santiago, D.N., Pabón-Saldaña, M., Darville, L., Fang, B., Rix, U., Yoder, S., Berglund, A., Koomen, J.M., Haura, E.B., and Abate-Daga, D.: 'An immunoproteomic approach to characterize the CAR interactome and signalosome', Science Signaling, 2019, 12, (568)
22  Altan-Bonnet, G., and Germain, R.N.: 'Modeling T Cell Antigen Discrimination Based on Feedback Control of Digital ERK Responses', PLoS Biology, 2005, 3, (11)
23  Condomines M, J., A., R., B., J., P., G., G., I., R., and M., S.: '466. Attenuated CTLA-4 Inhibition in CD19-Targeted T Cells Expressing a Second-Generation CD28-Based Chimeric Antigen Receptor', Molecular Therapy, 2012, 20
24  Tang, L., Zhang, Y., Hu, Y., Mei, H., and Li, Y.: 'T Cell Exhaustion and CAR-T Immunotherapy in Hematological Malignancies', BioMed Research International, 2021, 2021, pp. 1-8
25  Lynn, R.C., Weber, E.W., Sotillo, E., Gennert, D., Xu, P., Good, Z., Anbunathan, H., Lattin, J., Jones, R., Tieu, V., Nagaraja, S., Granja, J., de Bourcy, C.F.A., Majzner, R., Satpathy, A.T., Quake, S.R., Monje, M., Chang, H.Y., and Mackall, C.L.: 'c-Jun overexpression in CAR T cells induces exhaustion resistance', Nature, 2019, 576, (7786), pp. 293-300